\documentclass[twocolumn,showpacs,preprintnumbers,amsmath,amssymb]{revtex4}
\usepackage{amsmath}
\usepackage{bm}
\usepackage{graphicx}

\newcommand{\ket}[1]{|#1\rangle}

\renewcommand{\vec}[1]{\mathbf{#1}}
\newcommand{\unitv}[1]{\mathbf{\hat{#1}}}

\begin{document}

\title{A scalable Bose-Einstein condensate Sagnac interferometer 
in a linear trap}
\author{J. H. T. Burke and C. A. Sackett}
\affiliation{University of Virginia, Dept. of Physics, Charlottesville, VA 22904
, USA}
\date{\today}


\begin{abstract}
We demonstrate a two-dimensional atom interferometer in a
harmonic magnetic waveguide using a Bose-Einstein condensate. Such an 
interferometer could measure rotation using the Sagnac effect.  
Compared to free space interferometers, larger
interactions times and enclosed areas can in principle be achieved, since the 
atoms are
not in free fall.  In this implementation, we induce the atoms to 
oscillate along one 
direction
by displacing the trap center.  We then split and recombine the atoms along an orthogonal direction,
using an off-resonant optical standing wave.  We enclose a maximum 
effective area of 0.1~mm$^2$, limited by fluctuations in the
initial velocity and the coherence time of the
 interferometer.  We argue that this arrangement is 
scalable to enclose larger areas by increasing 
the coherence time and then making repeated loops. 
\end{abstract}

\pacs{03.75.Dg, 37.25.k}

\maketitle

Atom interferometry has proven useful for a variety of precision measurements,
notably including rotations
\cite{Berman97,Gustavson00,Durfee06}. 
Through the Sagnac effect \cite{Sagnac1913}, an 
interferometer that encloses area $A$ on a platform rotating at
rate $\Omega$ develops a phase proportional to the product of $A$ and $\Omega$.
Greater sensitivity is therefore obtained by increasing the area.
The best atom gyroscope at present
uses a thermal atomic beam in a 2-m-long Mach-Zehner configuration, with
an enclosed area of 30~mm$^2$ \cite{Durfee06}.  
It exhibits impressive performance, but the considerable length
of the device limits potential applications.
 
A possible resolution to this problem is guided-wave atom
interferometry, in which the atoms are continuously confined by
optical or magnetic fields.  The guiding fields can 
direct the atoms along more compact trajectories than possible in
free space.  For instance, an enclosed area comparable to the
above could be obtained by passing the atoms around a circular loop
of only 6-mm diameter.  In addition, the confining potential supports the
atoms against gravity, permitting longer interaction times.
A significant effort is underway to develop such devices
\cite{Sauer01,Wu04,Gupta05,Wu07,Jo07,Griffin08}, but to date, only 
relatively small enclosed areas have been demonstrated.
We present here a design that we argue has good potential to 
scale to large area and take 
advantage of the benefits of the guided-wave approach.

The design is based on a linear guided-wave interferometer
\cite{Garcia06}.  A Bose-Einstein condensate
is produced in a harmonic trap.  One axis ($y$) of the trap is
weakly confining, and an off-resonant standing wave laser passing along
that axis is used to split, manipulate, and recombine the atomic
wave packets.  The trajectory is shown in Fig.~1(a).  Note that
a reciprocal trajectory is used, in which both packets traverse 
identical paths.
This 
causes static perturbations from the confining potential to largely
cancel \cite{Burke08}.  To generate an enclosed area, we operate
the interferometer with atoms that are also moving in the transverse ($x$)
direction.  In that direction, the atoms undergo harmonic oscillation.
The laser pulses are timed so that a turning point in $x$
occurs at the midpoint of the interferometer, as shown in Fig.~1(b).
The trajectory in the $xy$-plane is shown in Fig.~1(c), and
is clearly both area-enclosing and reciprocal.

\begin{figure}
{\includegraphics[width=3.2in]{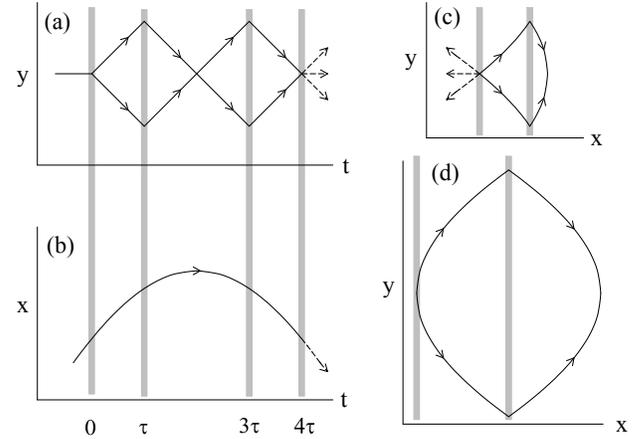}}
\caption{\label{fig:Tra} (a)--(c) Sagnac interferometer trajectory.
(a) Motion along the guide axis ($y$) vs time.  (b) Motion in the
direction ($x$) transverse to the guide.  (c) Trajectory in the $xy$-plane.
Gray bars indicate the times and positions of interactions with a standing
 wave laser beam that is parallel
 to the $y$-axis.  Solid curves show the packet trajectories and
dashed segments show the possible output states.
(d) Trajectory obtained when the packets begin at a turning point in $x$,
allowing for multiple loops.
All figures are shown with a consistent scale.
}
\end{figure}

Our technique is similar to that of Wu~{\em et al.}\ \cite{Wu07}, but there 
the transverse motion is achieved by translating the guide itself
rather than by excitation within in the guide.  The primary
difference here is our use of a Bose-Einstein condensate, compared
to laser-cooled atoms. 
This makes the moving guide approach challenging, since the guide
would need to be tightly confining to ensure the atoms followed it 
adiabatically.  For a condensate, tight confinement increases
interaction effects that can spoil the interference.

Using a condensate does, however, allow a higher degree of control.  
When applied to a thermal sample, the standing-wave laser pulses
produce many different interfering paths with differing enclosed areas.
These parallel loops produce a complicated output state, from which
a Sagnac signal must be reconstructed.  Wu~{\em et al.}\ demonstrate
how this can be achieved (see also \cite{Tonyushkin08}), 
but it is not yet clear whether their methods can be
extended to the accuracy needed for precision measurements.  
In addition, the large range of initial atomic velocities makes the
interferometer imperfectly reciprocal and thus more sensitive to 
errors.  In contrast, the low momentum spread of a condensate
permits our interferometer to be operated with a single reciprocal
trajectory to a high degree of accuracy.

The linear interferometer on which our method is based has been
described previously \cite{Garcia06}. 
To start, a Bose-Einstein condensate of
$N = 3\times10^{4}$ $^{87}$Rb atoms is prepared in the $F=2, m_{F}=2$
hyperfine state and held in a harmonic time-orbiting potential (TOP) trap.
Confinement is intentionally weak, with
atom oscillation frequencies  $(\omega_x, \omega_y, \omega_z)
\approx 2\pi\times(6.0, 1.1, 3.3)$~Hz.  Weak confinement in combination with 
support against gravity is achieved
by modulating the magnetic quadrupole field in phase with the
rotating bias field of the TOP trap \cite{Reeves05}.

We excite the
transverse motion by suddenly changing the phase difference between the
bias and quadrupole fields, which shifts the trap minimum along $x$.  
The displacement is not purely transverse,
so in general oscillations are induced in all three directions.  
However, at a suitable time later the phase is switched back, 
causing the $x$ oscillation to be enhanced while the $y$ oscillation 
is reduced.  Typical values of the oscillation amplitudes are
$C_x=1$~mm and residual amplitudes of $100~\mu$m in the other directions.

The motion along $y$ is controlled by a standing-wave laser 
at a wavelength $\lambda = 780.1$~nm, 
70~GHz blue of the $5S_{1/2} \leftrightarrow 5P_{3/2}$ laser-cooling 
transition.
The laser couples states of momentum $p_y =  2n\hbar k$ where $k = 2\pi/\lambda$ 
and $n$ is an integer.  The corresponding velocity $v_0 = 2\hbar k/m$
is 1.2~cm/s.
In particular, the interferometer uses 
the beam-splitting transition $\ket{0} \leftrightarrow (\ket{2\hbar
k}+\ket{{-2\hbar k}})/\sqrt{2}$ and the reflection transition
$\ket{2\hbar k}\leftrightarrow\ket{{-2\hbar k}}$, both of
which can be implemented with high precision \cite{Wu05b,Hughes07}.
We use a 1-cm diameter laser beam
to encompass the range of transverse locations needed for the pulses. 

To create the interferometer, a splitting pulse is applied at time
$t = 0$, reflections at times $t = \tau$ and $3\tau$, and a
splitting pulse again at $t = 4\tau$.  In general, the output
consists of three momentum states, $p_y = 0, \pm 2\hbar k$.
The populations $N_i$ of the states depends on the interferometer phase $\phi$,
with $N_0/N = (1+V\cos\phi)/2$ for visibility $V$.
We vary $\phi$ in a controlled way by adjusting the frequency
of the coupling laser before the final splitting operation,
which has the effect of shifting the standing wave along $y$.
This permits an interference curve $N_0(\phi)$ to be mapped out and
the visibility determined.
With no transverse excitation, interference is observed for total
times $4\tau$ up to 72~ms.  This is limited by phase gradients imposed
by the non-uniform potential along the guide axis \cite{Burke08}.
In the experiments described here, we operate with $\tau = 5$ and 10 ms,
for which the linear interferometer has a visibility of about 0.9.

The two-dimensional interferometer of Fig.~1(c) encloses area
\begin{equation}        
A \approx \frac{4v_0C_x}{\omega_x}(2\sin{\omega_x\tau}-\sin{2\omega_x\tau}),
\label{equ:area}
\end{equation} 
valid for $\omega_y\tau \ll 1$.  The effective area for a Sagnac
interferometer is twice this, since both packets complete a full
circuit of the loop.
We implemented the interferometer with oscillation amplitudes
$C_x=0.4$~mm, $0.8$~mm, and 1.3~mm.  Results for the visibility
are shown in Fig.~2.  The enclosed areas are too small here to observe
the Sagnac effect from the Earth's rotation.

\begin{figure}[t]
{\includegraphics[width=3.2in]{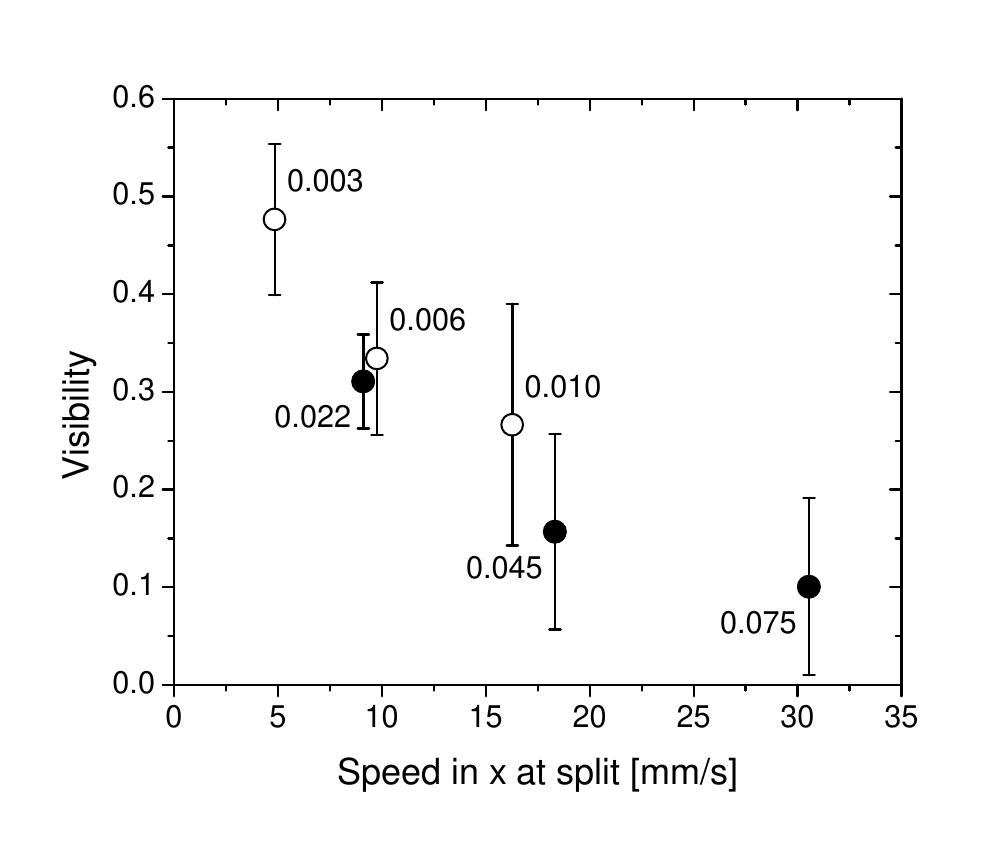}}
\caption{\label{fig:vis} Visibility of interferomete
as a function of the 
speed of the atoms at the time of the initial 
beam-splitting operation.  Open circles and filled 
circles represent $\tau$ equal 5~ms and 10~ms respectively. 
The labels denote the area enclosed in mm$^2$.}
\end{figure}

The data suggests that the interferometer coherence decays as the
the transverse velocity $v_x$ increases, 
with $v_x$ calculated as $C_x\omega_x \sin\omega_x\tau$.
The decay is due to phase noise.  
We observe that the rms variation in $N_0/N$
is about the same throughout, and corresponds to
an underlying visibility of about 0.5.
This suggests that the loss of visibility comes from to a noise effect
related to $v_x$.

Ideally, the interferometer operation is independent of $v_x$,
since the potential is separable.
If the motion in $x$ were identical for both packets, then any phase
accumulation due to that motion would be identical and thus cancel.
However, there are several ways that motion in $x$ can couple to the
interferometer direction $y$.  First, as noted above, driving the
$x$-oscillation does produce excitation in $y$ of about 0.6~mm/s
amplitude.  We minimized this effect by selecting the starting time
for the experiment such that the motion along $y$ was near an
extremum.  We estimate that the resulting $v_y$ was below 0.1~mm/s,
which is too small to explain the observed effects.

More seriously, the standing wave laser is not perfectly
aligned to the principle axis of the trap.
Observations of trajectories in the trap indicate alignment errors less
than a few degrees in the horizontal $(x)$ direction, and a larger
error of about $6^{\circ}$ in the vertical $(z)$ direction. 
Atomic motion along $x$ and $z$ therefore does produce a component
along the interferometer direction with magnitude similar to that of
the first effect.  
The beam misalignment is difficult to correct in the current configuration 
due to limited optical access and the large diameter of the standing-wave
beam.

An initial velocity parallel to the standing wave
can manifest itself in two ways.  First, the beam-splitting
and reflection operations are velocity-dependent \cite{Hughes07}.
Errors in the operations can leave atoms behind in unwanted momenta,
where they can affect the output phase.  For instance, if the
beam-splitting operation leaves a residual wave packet at $p = 0$, 
these atoms will continue through the interferometer and act as
an additional input state to the final recombination pulse,
changing the output in a phase sensitive way.  Since the interference
signal depends on the square root of the number of atoms in the stray
packet, even small errors in the standing-wave operations can result
in significant phase shifts.  A characteristic of this type of
error is a phase-dependent asymmetry between the $+2\hbar k$ and $-2\hbar k$
populations in the interferometer output, which we do observe at
larger $v_x$.

Even atoms in the correct motional states can acquire phase noise
through variations in their trajectory as they 
traverse the interferometer.
This effect can be calculated from the classical action.  In
a harmonic potential, the action is 
\begin{equation}
\Phi=\frac{S}{\hbar}=\frac{m}{2\hbar}\int dt\left(|\vec{v}|^2-|\vec{q}|^2\right)
\label{eq:phi}
\end{equation}
where $\vec{v}$ is the velocity, $\vec{x}$ is the position, 
and $q_i\equiv \omega_ix_i$.  The integral in (\ref{eq:phi}) can be
carried out for our trajectory, yielding a phase
\begin{equation}
\Phi = \frac{mv_0}{\hbar}\unitv{b}\cdot\vec{A}
\label{eq:Phi}
\end{equation}
where $\unitv{b}$
is the unit vector pointing in the direction of the Bragg beam and
the vector $\vec{A}$ is defined by
$A_i=C_if(\omega_i\tau)$ for
\begin{equation}
f(\omega\tau)  = \cos(6\omega\tau)-2\cos(5\omega\tau)
   +2\cos(3\omega\tau)-\cos(2\omega\tau) .
\label{eq:DPh}
\end{equation}
If $\omega\tau \ll 1$, $f \rightarrow 8(\omega\tau)^4$.

For example, using a horizontal alignment error of 1$^\circ$,
a vertical error of $6^\circ$, 
an amplitude $C_x = 1$~mm with $C_y = C_z = 0$, and $\tau=10$~ms,
we find $\Phi=12$~radians.  If
$C_x$ fluctuates with a standard deviation $\sigma=pC_x$, the visibility will 
be reduced by a 
factor $\exp{[-p^2\Phi^2/2]}$.  We observe fluctuations in the
$x$ amplitude of about 5\%, corresponding to a visibility decrease of
60\%, in qualitative agreement with the data above.

On the other hand, values of $\Phi$ obtained with $\tau = 5$~ms
are generally too small to explain the observed performance.  
The additional noise may derive from the degraded standing-wave operation
described above, since that effect is independent of $\tau$ apart from the
velocity's own dependence on $\tau$.
Modeling suggests that velocity errors on the order of those observed
would be sufficient \cite{Hughes07}.

We also estimate the effect of a small 
anharmonic term in the guide potential, $m\alpha x^3/3$.
Again, the resulting phase is calculated using the action, now with
the approximate trajectory for an anharmonic oscillator
\cite{Landau60}.  To leading order, we find 
\begin{equation}
\Phi_{\alpha}\approx 
\frac{28}{3}\frac{mv_0}{\hbar}(\alpha\tau^2)\sum_i{b_iC_i^2.
}
\end{equation}
Numerical modeling of the trap current conductors suggests
$\alpha \approx 10^3$~m$^{-1}$s$^{-2}$.  With
$C_x=1$~mm, this yields $\Phi_{\alpha}\approx 0.25$~rad.
Fluctuations on this value will be too small to contribute 
significantly to the observed noise, but the effect is not
negligible.

In principle, it is possible to model all of these noise effects together
and attempt to reproduce the behavior seen in Fig.~2.  We pursued
such an approach, but found that the results to be very sensitive
to the motional noise amplitudes and the alignment angle errors.
Our experimental knowledge of these parameters is insufficient to
constrain the model enough to be useful, in that the experimental
behavior could be reproduced for many different sets of error parameters.
The model does, however, further support the conclusion that the
mechanisms discussed are sufficient to explain the experimental performance.
   
Despite the fact that the performance is limited, 
we believe our approach has
promise for creating a compact guided-wave gyroscope.  
A significant advantage
is that the technique is continuously scalable from zero area, a 
feature that will continue to make troubleshooting easier as the area 
increases. 
Additionally, the technique is in principle 
capable of enclosing larger areas by 
making repeated loops. 
If it can be operated 
with $\tau=\pi/(2\omega_x)\approx40$~ms, then the split pulse 
occurs at a turning point in $x$.  The trajectory then resembles a vesica 
piscis as seen in Fig. \ref{fig:Tra}(d).  The area in this limit is 
$A=8C_x v_0/\omega_x$, and since the trajectory is closed, 
the atoms can complete multiple orbits.
Furthermore, in the configuration the $x$ component of the velocity is
near zero at the time of the beam-splitting pulses, which would significantly
reduce the sensitivity to beam angle described above.

We are currently installing a new magnetic trap apparatus with a 
more uniform potential along $y$, avoiding the problem
of longitudinal decoherence.  
The new apparatus will also provide optical access allowing the 
beam alignment errors to be more easily remedied.  
With these improvements, we
expect to increase the
useable interferometer duration and reach $\tau = \pi/2\omega_x$.
We have previously observed interference at 
one-second measurement times \cite{Burke08}, and if that can be
attained here, up to five orbits through the trap would be achievable.
This would enclose an effective area of 26~mm$^2$ for $C_x=1$~mm,
comparable to present free-atom gyroscopes, but taking up much less 
space.  It might also be possible to apply the ideas of 
\cite{Search09} to this geometry, to increase the sensitivity even further.

Our method is similar to an approach using atoms confined in
a cylindrically symmetric harmonic trap  \cite{Horikoshi07}.  
In that case, reflection operations are not necessary, since the
potential itself guides the packets in a circular orbit.
After a complete orbit, both the
$x$ and $y$ terms in Eq.~(\ref{eq:Phi}) vanish,
leaving 
\begin{equation}
\Phi=\frac{2mv_0}{\hbar}A_z\sin{(\epsilon)}\sin^2{\left(\frac{2\pi\omega_z}{\omega_{xy}}\right)}.
\end{equation}  
This too will vanish if
$\omega_z$ is equal to an integer multiple of $\omega_{xy}/2$.  
We are presently designing a trap to test this symmetric approach as well.

In either geometry, stray interference with erroneous paths may
still be problematic, since the beam-splitter operations will never be perfect.
The dominant error, from atoms left with $p =0$ after the split, could
be resolved by removing those atoms from the trap during the interferometer.
In the linear trap, this would require a focused laser beam, but in the
cylindrical trap, an rf-evaporation pulse tuned to the bottom of the trap
potential would suffice.

We compare these results with other guided-atom gyroscope efforts. Several 
experiments have demonstrated ring-shaped guides
\cite{Sauer01,Griffin08,Gupta05,Wu04}, but none as yet have exhibited 
interference.  Besides the practical problems in creating these potentials such 
as scalability, there is a more fundamental problem. A ring-guide
system will generally 
exhibit a phase linearly proportional to the initial velocity of the atoms,
since tangential motion of atoms around the ring is equivalent to a rotational 
and thus subject to the Sagnac effect.  
This effect is much larger than that of Eq.~(\ref{eq:Phi}):
for an equivalent area of $0.05$~mm$^2$, a ring interferometer would
exhibit phase noise of 1~rad for a velocity fluctuation of only
1~$\mu$m/s.  In either of the geometries discussed here, this
phase largely cancels due to reflection by either the standing wave
or the harmonic potential.

The comparable experiment of \cite{Wu07} avoids this problem, and
the use of thermal atoms provides the opportunity to average over
unwanted interfering paths.
However, the inefficiency of the beam-splitting and reflection operations
will make it difficult to achieve trajectories with multiple orbits.
Additionally, thermal expansion of the sample disrupts 
the reciprocality of the trajectories as atoms in the sample move 
relative to the trap center during the measurement.  
The use of condensate atoms may thus offer several long-term advantages.

We have demonstrated how a linear interferometer of ultracold atoms in a weak 
guiding potential can be extended to perform gyroscopic measurements.  This 
gyroscope has an effective enclosed area of 0.05 mm$^2$, but shows promise of 
being scalable to larger area.  
Current performance is limited by initial velocity fluctuations, but
stabilization of the oscillation-inducing method, 
improvements in optical access, and optimization of the 
trap geometry should provide large gains.  We hope that in the near 
future precision rotation measurements with this or a similar device will be 
possible. 

We thank D.~Stamper-Kurn for useful conversations, and B.~Deissler and
K.J.~Hughes for their early work on the project.  
This work was supported by the 
Defense Advanced Research Projects
Agency (Grant No. 51925-PH-DRP) and by the National
Science Foundation (Grant No. PHY-0244871).


\begin{thebibliography}{19}
\expandafter\ifx\csname natexlab\endcsname\relax\def\natexlab#1{#1}\fi
\expandafter\ifx\csname bibnamefont\endcsname\relax
  \def\bibnamefont#1{#1}\fi
\expandafter\ifx\csname bibfnamefont\endcsname\relax
  \def\bibfnamefont#1{#1}\fi
\expandafter\ifx\csname citenamefont\endcsname\relax
  \def\citenamefont#1{#1}\fi
\expandafter\ifx\csname url\endcsname\relax
  \def\url#1{\texttt{#1}}\fi
\expandafter\ifx\csname urlprefix\endcsname\relax\def\urlprefix{URL }\fi
\providecommand{\bibinfo}[2]{#2}
\providecommand{\eprint}[2][]{\url{#2}}

\bibitem[{\citenamefont{Berman}(1997)}]{Berman97}
\bibinfo{editor}{\bibfnamefont{P.~R.} \bibnamefont{Berman}}, ed.,
  \emph{\bibinfo{title}{Atom Interferometry}} (\bibinfo{publisher}{Academic
  Press}, \bibinfo{address}{San Diego}, \bibinfo{year}{1997}).

\bibitem[{\citenamefont{Gustavson et~al.}(2000)\citenamefont{Gustavson,
  Landragin, and Kasevich}}]{Gustavson00}
\bibinfo{author}{\bibfnamefont{T.~L.} \bibnamefont{Gustavson}},
  \bibinfo{author}{\bibfnamefont{A.}~\bibnamefont{Landragin}},
  \bibnamefont{and} \bibinfo{author}{\bibfnamefont{M.~A.}
  \bibnamefont{Kasevich}}, \bibinfo{journal}{Class. Quantum Grav.}
  \textbf{\bibinfo{volume}{17}}, \bibinfo{pages}{2385} (\bibinfo{year}{2000}).

\bibitem[{\citenamefont{Durfee et~al.}(2006)\citenamefont{Durfee, Shaham, and
  Kasevich}}]{Durfee06}
\bibinfo{author}{\bibfnamefont{D.}~\bibnamefont{Durfee}},
  \bibinfo{author}{\bibfnamefont{Y.}~\bibnamefont{Shaham}}, \bibnamefont{and}
  \bibinfo{author}{\bibfnamefont{M.}~\bibnamefont{Kasevich}},
  \bibinfo{journal}{Phys. Rev. Lett.} \textbf{\bibinfo{volume}{97}},
  \bibinfo{pages}{240801} (\bibinfo{year}{2006}).

\bibitem[{\citenamefont{Sagnac}(1913)}]{Sagnac1913}
\bibinfo{author}{\bibfnamefont{G.}~\bibnamefont{Sagnac}}, \bibinfo{journal}{C.
  R. Acad. Sci.} \textbf{\bibinfo{volume}{95}}, \bibinfo{pages}{708}
  (\bibinfo{year}{1913}).

\bibitem[{\citenamefont{Sauer et~al.}(2001)\citenamefont{Sauer, Barrett, and
  Chapman}}]{Sauer01}
\bibinfo{author}{\bibfnamefont{J.~A.} \bibnamefont{Sauer}},
  \bibinfo{author}{\bibfnamefont{M.~D.} \bibnamefont{Barrett}},
  \bibnamefont{and} \bibinfo{author}{\bibfnamefont{M.~S.}
  \bibnamefont{Chapman}}, \bibinfo{journal}{Phys. Rev. Lett.}
  \textbf{\bibinfo{volume}{87}}, \bibinfo{pages}{270401}
  (\bibinfo{year}{2001}).

\bibitem[{\citenamefont{Wu et~al.}(2004)\citenamefont{Wu, Rooijakkers, Striehl,
  and Prentiss}}]{Wu04}
\bibinfo{author}{\bibfnamefont{S.}~\bibnamefont{Wu}},
  \bibinfo{author}{\bibfnamefont{W.}~\bibnamefont{Rooijakkers}},
  \bibinfo{author}{\bibfnamefont{P.}~\bibnamefont{Striehl}}, \bibnamefont{and}
  \bibinfo{author}{\bibfnamefont{M.}~\bibnamefont{Prentiss}},
  \bibinfo{journal}{Phys. Rev. A} \textbf{\bibinfo{volume}{70}},
  \bibinfo{pages}{013409} (\bibinfo{year}{2004}).

\bibitem[{\citenamefont{Gupta et~al.}(2005)\citenamefont{Gupta, Murch, Moore,
  Purdy, and Stamper-Kurn}}]{Gupta05}
\bibinfo{author}{\bibfnamefont{S.}~\bibnamefont{Gupta}},
  \bibinfo{author}{\bibfnamefont{K.~W.} \bibnamefont{Murch}},
  \bibinfo{author}{\bibfnamefont{K.~L.} \bibnamefont{Moore}},
  \bibinfo{author}{\bibfnamefont{T.~P.} \bibnamefont{Purdy}}, \bibnamefont{and}
  \bibinfo{author}{\bibfnamefont{D.~M.} \bibnamefont{Stamper-Kurn}},
  \bibinfo{journal}{Phys. Rev. Lett.} \textbf{\bibinfo{volume}{95}},
  \bibinfo{pages}{143201} (\bibinfo{year}{2005}).

\bibitem[{\citenamefont{Wu et~al.}(2007)\citenamefont{Wu, Su, and
  Prentiss}}]{Wu07}
\bibinfo{author}{\bibfnamefont{S.}~\bibnamefont{Wu}},
  \bibinfo{author}{\bibfnamefont{E.}~\bibnamefont{Su}}, \bibnamefont{and}
  \bibinfo{author}{\bibfnamefont{M.}~\bibnamefont{Prentiss}},
  \bibinfo{journal}{Phys. Rev. Lett.} \textbf{\bibinfo{volume}{99}},
  \bibinfo{pages}{173201} (\bibinfo{year}{2007}).

\bibitem[{\citenamefont{Jo et~al.}(2007)\citenamefont{Jo, Shin, Will, Pasquini,
  Saba, Ketterle, and Pritchard}}]{Jo07}
\bibinfo{author}{\bibfnamefont{G.-B.} \bibnamefont{Jo}},
  \bibinfo{author}{\bibfnamefont{Y.}~\bibnamefont{Shin}},
  \bibinfo{author}{\bibfnamefont{S.}~\bibnamefont{Will}},
  \bibinfo{author}{\bibfnamefont{T.}~\bibnamefont{Pasquini}},
  \bibinfo{author}{\bibfnamefont{M.}~\bibnamefont{Saba}},
  \bibinfo{author}{\bibfnamefont{W.}~\bibnamefont{Ketterle}}, \bibnamefont{and}
  \bibinfo{author}{\bibfnamefont{D.}~\bibnamefont{Pritchard}},
  \bibinfo{journal}{Phys. Rev. Lett.} \textbf{\bibinfo{volume}{98}},
  \bibinfo{pages}{030407} (\bibinfo{year}{2007}).

\bibitem[{\citenamefont{Griffin et~al.}(2008)\citenamefont{Griffin, Riis, and
  Arnold}}]{Griffin08}
\bibinfo{author}{\bibfnamefont{P.~F.} \bibnamefont{Griffin}},
  \bibinfo{author}{\bibfnamefont{E.}~\bibnamefont{Riis}}, \bibnamefont{and}
  \bibinfo{author}{\bibfnamefont{A.~S.} \bibnamefont{Arnold}},
  \bibinfo{journal}{Phys. Rev. A} \textbf{\bibinfo{volume}{77}},
  \bibinfo{pages}{051402(R)} (\bibinfo{year}{2008}).

\bibitem[{\citenamefont{Garcia et~al.}(2006)\citenamefont{Garcia, Deissler,
  Hughes, Reeves, and Sackett}}]{Garcia06}
\bibinfo{author}{\bibfnamefont{O.}~\bibnamefont{Garcia}},
  \bibinfo{author}{\bibfnamefont{B.}~\bibnamefont{Deissler}},
  \bibinfo{author}{\bibfnamefont{K.~J.} \bibnamefont{Hughes}},
  \bibinfo{author}{\bibfnamefont{J.~M.} \bibnamefont{Reeves}},
  \bibnamefont{and} \bibinfo{author}{\bibfnamefont{C.~A.}
  \bibnamefont{Sackett}}, \bibinfo{journal}{Phys. Rev. A}
  \textbf{\bibinfo{volume}{74}}, \bibinfo{pages}{031601(R)}
  (\bibinfo{year}{2006}).

\bibitem[{\citenamefont{Burke et~al.}(2008)\citenamefont{Burke, Deissler,
  Hughes, and Sackett}}]{Burke08}
\bibinfo{author}{\bibfnamefont{J.~H.~T.} \bibnamefont{Burke}},
  \bibinfo{author}{\bibfnamefont{B.}~\bibnamefont{Deissler}},
  \bibinfo{author}{\bibfnamefont{K.~J.} \bibnamefont{Hughes}},
  \bibnamefont{and} \bibinfo{author}{\bibfnamefont{C.~A.}
  \bibnamefont{Sackett}}, \bibinfo{journal}{Phys. Rev. A.}
  \textbf{\bibinfo{volume}{78}}, \bibinfo{pages}{043404}
  (\bibinfo{year}{2008}).

\bibitem[{\citenamefont{Tonyushkin and Prentiss}(2008)}]{Tonyushkin08}
\bibinfo{author}{\bibfnamefont{A.}~\bibnamefont{Tonyushkin}} \bibnamefont{and}
  \bibinfo{author}{\bibfnamefont{M.}~\bibnamefont{Prentiss}},
  \bibinfo{journal}{Phys. Rev. A} \textbf{\bibinfo{volume}{78}},
  \bibinfo{pages}{053625} (\bibinfo{year}{2008}).

\bibitem[{\citenamefont{Reeves et~al.}(2005)\citenamefont{Reeves, Garcia,
  Deissler, Baranowski, Hughes, and Sackett}}]{Reeves05}
\bibinfo{author}{\bibfnamefont{J.~M.} \bibnamefont{Reeves}},
  \bibinfo{author}{\bibfnamefont{O.}~\bibnamefont{Garcia}},
  \bibinfo{author}{\bibfnamefont{B.}~\bibnamefont{Deissler}},
  \bibinfo{author}{\bibfnamefont{K.~L.} \bibnamefont{Baranowski}},
  \bibinfo{author}{\bibfnamefont{K.~J.} \bibnamefont{Hughes}},
  \bibnamefont{and} \bibinfo{author}{\bibfnamefont{C.~A.}
  \bibnamefont{Sackett}}, \bibinfo{journal}{Phys. Rev. A}
  \textbf{\bibinfo{volume}{72}}, \bibinfo{pages}{051605(R)}
  (\bibinfo{year}{2005}).

\bibitem[{\citenamefont{Wu et~al.}(2005)\citenamefont{Wu, Wang, Diot, and
  Prentiss}}]{Wu05b}
\bibinfo{author}{\bibfnamefont{S.}~\bibnamefont{Wu}},
  \bibinfo{author}{\bibfnamefont{Y.}~\bibnamefont{Wang}},
  \bibinfo{author}{\bibfnamefont{Q.}~\bibnamefont{Diot}}, \bibnamefont{and}
  \bibinfo{author}{\bibfnamefont{M.}~\bibnamefont{Prentiss}},
  \bibinfo{journal}{Phys. Rev. A} \textbf{\bibinfo{volume}{71}},
  \bibinfo{pages}{043602} (\bibinfo{year}{2005}).

\bibitem[{\citenamefont{Hughes et~al.}(2007)\citenamefont{Hughes, Deissler,
  Burke, and Sackett}}]{Hughes07}
\bibinfo{author}{\bibfnamefont{K.~J.} \bibnamefont{Hughes}},
  \bibinfo{author}{\bibfnamefont{B.}~\bibnamefont{Deissler}},
  \bibinfo{author}{\bibfnamefont{J.~H.~T.} \bibnamefont{Burke}},
  \bibnamefont{and} \bibinfo{author}{\bibfnamefont{C.~A.}
  \bibnamefont{Sackett}}, \bibinfo{journal}{Phys. Rev. A}
  \textbf{\bibinfo{volume}{76}}, \bibinfo{pages}{035601}
  (\bibinfo{year}{2007}).

\bibitem[{\citenamefont{Landau and Lifshitz}(1960)}]{Landau60}
\bibinfo{author}{\bibfnamefont{L.~D.} \bibnamefont{Landau}} \bibnamefont{and}
  \bibinfo{author}{\bibfnamefont{E.~M.} \bibnamefont{Lifshitz}},
  \emph{\bibinfo{title}{Mechanics}} (\bibinfo{publisher}{Pergamon},
  \bibinfo{address}{New York}, \bibinfo{year}{1960}), \bibinfo{edition}{3rd}
  ed.

\bibitem[{\citenamefont{Search et~al.}(2009)\citenamefont{Search, Toland, and
  Zivkovic}}]{Search09}
\bibinfo{author}{\bibfnamefont{C.}~\bibnamefont{Search}},
  \bibinfo{author}{\bibfnamefont{J.}~\bibnamefont{Toland}}, \bibnamefont{and}
  \bibinfo{author}{\bibfnamefont{M.}~\bibnamefont{Zivkovic}},
  \bibinfo{journal}{Phys. Rev. A} \textbf{\bibinfo{volume}{79}},
  \bibinfo{pages}{053607} (\bibinfo{year}{2009}).

\bibitem[{\citenamefont{Horikoshi and Nakagawa}(2007)}]{Horikoshi07}
\bibinfo{author}{\bibfnamefont{M.}~\bibnamefont{Horikoshi}} \bibnamefont{and}
  \bibinfo{author}{\bibfnamefont{K.}~\bibnamefont{Nakagawa}},
  \bibinfo{journal}{Phys. Rev. Lett.} \textbf{\bibinfo{volume}{99}},
  \bibinfo{pages}{180401} (\bibinfo{year}{2007}).

\end{thebibliography}

\end{document}